\documentclass[12pt]{article}

\usepackage{graphicx}

\begin{document}

\begin{center}
{\bf Acceleration of Universe by Nonlinear Electromagnetic Fields} \\
\vspace{5mm} S. I. Kruglov
\footnote{E-mail: serguei.krouglov@utoronto.ca}

\vspace{3mm}
\textit{Department of Chemical and Physical Sciences, University of Toronto,\\
3359 Mississauga Road North, Mississauga, Ontario L5L 1C6, Canada} \\
\vspace{5mm}
\end{center}

\begin{abstract}
A new model of nonlinear electromagnetic fields possessing a dimensional parameter $\beta$ is proposed. Electromagnetic fields are considered as the source of the gravitation field and accelerated expansion of the universe is driven by nonlinear electromagnetic fields. We consider the magnetic universe and the stochastic magnetic field is a source of the universe acceleration. After the universe inflation and the accelerated expansion the universe decelerates. We show the causality of the model and a classical stability at the deceleration phase. The spectral index, the tensor-to-scalar ratio, and the running of the spectral index were estimated that approximately fulfil the PLANK, WMAP, and BICEP2 data.
\end{abstract}

\section{Introduction}

The acceleration of the universe is proved by experimental data using the redshift of type Ia supernovae
and the cosmic microwave background (CMB). There are different ways to explain the universe acceleration: (1) to introduce the cosmological constant, $\Lambda$, in Einstein's equation; (2) to use a scalar field (an inflation field) in the matter Lagrangian with some potential function; (3) to modify the gravity theory by the replacement of the Ricci scalar $R$ in the Einstein-Hilbert action by the proper function $F(R)$; and there are other approaches.
In the first case it is not clear how to explain the smallness of the $\Lambda$ compared to the vacuum energy.
The equation of state in this case is $p=-\rho$ ($p$, $\rho$ are the pressure and the energy density, respectively) and the fluid is the dark energy. For the second case the introduction of the potential for the scalar field is not unique and the nature of the scalar field is unclear. In the third case there are many gravity models with different functions $F(R)$  \cite{Capozziello}. In this paper we use a new model of nonlinear electrodynamics (NLED), and electromagnetic fields are a source of gravity
that can drive the universe to accelerate. Thus, we do not modify General Relativity (GR) and such an approach can explain early time inflation. Just after Big Bang the electromagnetic and gravitational fields are very strong and quantum corrections should be taken into account \cite{Jackson}.
We imply that nonlinear electrodynamics suggested is an effective model of electromagnetic fields which is valid for strong fields and at weak fields it becomes Maxwell's electrodynamics. The NLED models coupled to the gravitation field can describe inflation \cite{Garcia}, \cite{Camara}, \cite{Elizalde}, \cite{Novello3}, \cite{Novello}, \cite{Novello1} and may produce negative pressure that drives the acceleration of the universe. The cosmological consequences of various NLED Lagrangians have previously been investigated in the papers \cite{Novello3}, \cite{Novello}, \cite{Novello1}, \cite{Vollick}, \cite{Salcedo}. Thus, nonlinear electrodynamics that we propose is new and has not been presented before.

There are related theoretical topics including foundations of electromagnetism. Within special theory of relativity the Lagrangian density for electromagnetic fields should be the function of two Lorentz invariants ${\cal F}=(\textbf{B}^2-\textbf{E}^2)/2$ and ${\cal G}= (\textbf{E}\cdot \textbf{B})$. The Lorentz invariant ${\cal G}$ is not a $P$-invariant and, therefore, to have the theory which is invariant under the inversion of coordinates the Lagrangian density should include the term ${\cal G}^2$. The extended Maxwell Lagrangian density ${\cal L}=-{\cal F} +4a{\cal F}^2+b{\cal G}^2$ with two parameters $a$ and $b$ was proposed in \cite{Kruglov}. The necessary and sufficient conditions to have no birefringence is $4a=b$. The Heisenberg-Euler Lagrangian density \cite{Heisenberg} due to one loop (quantum) corrections to classical electrodynamics gives the values $a=2\alpha^2/(45m^4)$ ($m$ is the mass of an electron), $b=7a$. The Lagrangian density with four parameters including the coupling with a pseudoscalar field and also with nonvanishing photon mass was considered in \cite{Ni}.
Born and Infeld \cite{Born} proposed the Lagrangian density for the electromagnetic field that smoothing singularity of point-like charges and gives the finite value of self-energy. The Born-Infeld electrodynamics like Maxwell's electrodynamics does not admit the effect of birefringence in the theory.
Some models of non-linear electrodynamics with functions of ${\cal F}$ and ${\cal G}$ in the Lagrangian density were investigated in \cite{Soleng}, \cite{Dymnikova}, \cite{Kruglov4}, \cite{Hendi}, \cite{Kruglov1}, \cite{Kruglov2}, \cite{Kruglov3}. Models for Lorentz violation in electrodynamics were considered in \cite{Kostelecky}, \cite{Kruglov5}.

For testing the foundations of classical electrodynamics in flat spacetime some experiments were arranged. There are three experiments for measuring vacuum birefringence, the BMV (Bir\'{e}fringence Magn\'{e}tique du Vide) experiment \cite{Battesti}, \cite{Rizzo}, the PVLAS (Polarizzazione del Vuoto con LASer) experiment \cite{Zavattini}, \cite{Valle} and the QA (QED vacuum birefringence and Axion search) experiment \cite{Chen}, \cite{Mei}.
The experiment for measuring the parameters of nonlinear electrodynamics of vacuum with laser interferometer techniques was also described in \cite{Denisov} and \cite{Ni}.

The structure of the paper is as follows. In section 2 we introduce a new model of nonlinear electromagnetic fields with a dimensional parameter $\beta$. The energy-momentum tensor is calculated possessing non-zero trace. The field equations are represented in the form of Maxwell's equations with the electric permittivity and magnetic permeability depending on the electromagnetic fields. The cosmology model with NLED fields coupled to gravity is investigated in Sec. 3. We consider the universe filled by stochastic magnetic fields. The dependance of the magnetic field on the scale factor is obtained and it is shown that there are no singularities of the energy density, pressure and the Ricci scalar. We demonstrate the universe acceleration when the scale factor is less than the critical value, and the critical scale factor is evaluated. In Sec. 4 the universe evolution is studied. We show the causality of the model and a classical stability at the deceleration phase. The Friedmann equation is solved and we find the dependance of the scale factor on time. In Sec. 5 we estimate the spectral index $n_s$, the tensor-to-scalar ratio $r$, and the running of the spectral index $\alpha_s$ which are in agreement with the PLANK, WMAP, and BICEP2 data. Conclusion is made in Sec. 6.

The units with $c=\hbar=\varepsilon_0=\mu_0=1$ and the metric $\eta=\mbox{diag}(-,+,+,+)$ are used.

\section{The model of nonlinear electromagnetic fields}

Let us consider the nonlinear electrodynamics with the Lagrangian density
\begin{equation}
{\cal L} = -\frac{1}{\beta}\arctan(\beta{\cal F}),
\label{1}
\end{equation}
with $\beta{\cal F}$ being dimensionless, ${\cal F}=(1/4)F_{\mu\nu}F^{\mu\nu}=(\textbf{B}^2-\textbf{E}^2)/2$, where $F_{\mu\nu}$ is the field strength tensor. The symmetric energy-momentum tensor can be obtained by varying the
action with respect to the metric \cite{Birula}
\begin{equation}
T^{\mu\nu}=H^{\mu\lambda}F^\nu_{~\lambda}-g^{\mu\nu}{\cal L},
\label{2}
\end{equation}
where
\begin{equation}
H^{\mu\lambda}=\frac{\partial {\cal L}}{\partial F_{\mu\lambda}}=\frac{\partial {\cal L}}{{\partial\cal F}}F^{\mu\lambda}=-\frac{F^{\mu\lambda}}{1+\left(\beta {\cal F}\right)^2}.
\label{3}
\end{equation}
Replacing Eq. (3) into (2) we find the symmetric energy-momentum tensor
\begin{equation}
T^{\mu\nu}=-\frac{F^{\mu\lambda}F^\nu_{~\lambda}}{1+\left(\beta {\cal F}\right)^2}-g^{\mu\nu}{\cal L},
\label{4}
\end{equation}
which has non-vanishing trace
\begin{equation}
{\cal T}\equiv T_{\mu}^{~\mu}=\frac{4}{\beta}\arctan(\beta{\cal F})- \frac{4{\cal F}}{1+\left(\beta {\cal F}\right)^2}.
\label{5}
\end{equation}
If $\beta\rightarrow 0$ one comes from Eq. (1) to classical electrodynamics, ${\cal L}\rightarrow -{\cal F}$ and trace (5) becomes zero, ${\cal T}\rightarrow 0$. Because the energy-momentum tensor trace is not zero the scale invariance is broken. This is the result of the introduction of the dimensional parameter $\beta$. Thus, the dilatation current is $D_\mu=x_\nu T_{\mu}^{~\nu}$, so that the divergence is $\partial_\mu D^\mu={\cal T}$.
The electric displacement field is given by $\textbf{D}=\partial{\cal L}/\partial \textbf{E}$, and according to Eq. (1) becomes
\begin{equation}
\textbf{D}=\frac{\textbf{E}}{1+\left(\beta {\cal F}\right)^2},
\label{6}
\end{equation}
so that $\textbf{D}=\varepsilon \textbf{E}$. As a result, the electric permittivity is given by
\begin{equation}
\varepsilon=\frac{1}{1+\left(\beta {\cal F}\right)^2}.
\label{7}
\end{equation}
The magnetic field can be obtained from the relation $\textbf{H}=-\partial{\cal L}/\partial \textbf{B}$, and is as follows:
\begin{equation}
\textbf{H}= \frac{\textbf{B}}{1+\left(\beta {\cal F}\right)^2},
\label{8}
\end{equation}
$\textbf{B}=\mu \textbf{H}$, and the magnetic permeability is $\mu=1/\varepsilon$. From Eqs. (6),(8) we find $\textbf{D}\cdot\textbf{H}=\varepsilon^2\textbf{E}\cdot\textbf{B}$. Because $\textbf{D}\cdot\textbf{H}\neq\textbf{E}\cdot\textbf{B}$ the dual symmetry is violated \cite{Gibbons} in this model.
Field equations follow from the Lagrangian density (1) and can be written, by virtue of Eqs. (6),(8), in the form of the first pair of Maxwell's equations
\begin{equation}
\nabla\cdot \textbf{D}= 0,~~~~ \frac{\partial\textbf{D}}{\partial
t}-\nabla\times\textbf{H}=0.
\label{9}
\end{equation}
From the Bianchi identity $\partial_\mu \widetilde{F}_{\mu\nu}=0$ ($\widetilde{F}_{\mu\nu}$ is a dual tensor), one finds the second pair of the Maxwell equations
\begin{equation}
\nabla\cdot \textbf{B}= 0,~~~~ \frac{\partial\textbf{B}}{\partial
t}+\nabla\times\textbf{E}=0.
\label{10}
\end{equation}
As the electric permittivity $\varepsilon$, Eq. (7), and the magnetic permittivity, $\mu=1/\varepsilon$, depend on the fields $\textbf{E}$, $\textbf{B}$, Eqs. (6), (8), (9), (10) represent nonlinear Maxwell's equations.

\section{The cosmology model}

In cosmology electromagnetic fields play an important role because of the CMB observation.
We consider the effective theory of electromagnetic fields described in the previous section coupled with the gravitation fields. The action of GR coupled with the nonlinear electromagnetic field (1) is given by
\begin{equation}
S=\int d^4x\sqrt{-g}\left[\frac{1}{2\kappa^2}R+ {\cal L}\right],
\label{11}
\end{equation}
where $\kappa^{-1}=M_{Pl}$, $M_{Pl}$ is the reduced Planck mass, and $R$ is the Ricci scalar. NLED is only a source of gravity here. Varying action (11) we obtain the Einstein and electromagnetic field equations
\begin{equation}
R_{\mu\nu}-\frac{1}{2}g_{\mu\nu}R=-\kappa^2T_{\mu\nu},
\label{12}
\end{equation}
\begin{equation}
\partial_\mu\left(\frac{\sqrt{-g}F^{\mu\nu}}{(\beta{\cal F})^2+1}\right)=0.
\label{13}
\end{equation}
In the absence of gravity Eq. (13) is equivalent to the system of Maxwell's equations (6), (8), (9), (10).
Let us consider homogeneous and isotropic cosmological spacetime with the metric
\begin{equation}
ds^2=-dt^2+a(t)^2\left(dx^2+dy^2+dz^2\right),
\label{14}
\end{equation}
where $a(t)$ is a scale factor.
The electromagnetic fields play the role of the cosmic stochastic background. We imply that the wavelength of electromagnetic waves is smaller as compared to the curvature.
One can make the average of the electromagnetic fields that are sources in GR \cite{Tolman} producing the isotropy of the Friedman-Robertson-Walker (FRW) spacetime.
The electromagnetic fields averaged have the properties
\[
<\textbf{E}>=<\textbf{B}>=0,~~~~<E_iB_j>=0,
\]
\begin{equation}
<E_iE_j>=\frac{1}{3}E^2g_{ij},~~~~<B_iB_j>=\frac{1}{3}B^2g_{ij},
\label{15}
\end{equation}
where the averaging brackets $<>$ mean an average over a volume larger than the radiation
wavelength and smaller compared to the spacetime curvature. In the following the brackets $<>$ will be omitted for simplicity. With the conditions (15) the energy-momentum tensor of NLED can be represented as a perfect fluid \cite{Novello1}.
We obtain the energy density $\rho$ and the pressure $p$ corresponding to the energy-momentum tensor (4)
\begin{equation}
\rho=-{\cal L}-E^2\frac{\partial{\cal L}}{{\cal \partial F}}=\frac{E^2}{\left(\beta{\cal F}\right)^2+1}+\frac{1}{\beta}\arctan(\beta{\cal F}),
\label{16}
\end{equation}
\begin{equation}
p={\cal L}-\frac{2B^2-E^2}{3}\frac{\partial{\cal L}}{{\cal \partial F}}=\frac{2B^2-E^2}{3\left[(\beta{\cal F})^2+1\right]}-\frac{1}{\beta}\arctan(\beta{\cal F}).
\label{17}
\end{equation}
Using FRW metric (14) and Einstein's equation (12), one finds the Friedmann equation
\begin{equation}
3\frac{\ddot{a}}{a}=-\frac{\kappa^2}{2}\left(\rho+3p\right),
\label{18}
\end{equation}
where dots over the $a$ mean the derivatives with respect to the cosmic time.
An accelerated expansion of the universe occurs if $\rho + 3p < 0$. Because the magnetic field lines are
disconnected \cite{Lemoine} only the magnetic field is important in cosmology. The electric field is screened due to the charged primordial plasma, and therefore, we investigate the case $E = 0$.
In the standard cosmological models there is a symmetry in the direction and, therefore, $<B_i> = 0$.
From Eqs. (16),(17) one can obtain
\begin{equation}
\rho+3p=\frac{2B^2}{1+(\beta^2 B^4)/4}-\frac{2}{\beta}\arctan\left(\frac{\beta B^2}{2}\right).
\label{19}
\end{equation}
The plot of the function $\beta(\rho+3p)$ versus $\beta B^2/2$ is given in Fig. 1.
\begin{figure}[h]
\includegraphics[height=4.0in,width=4.0in]{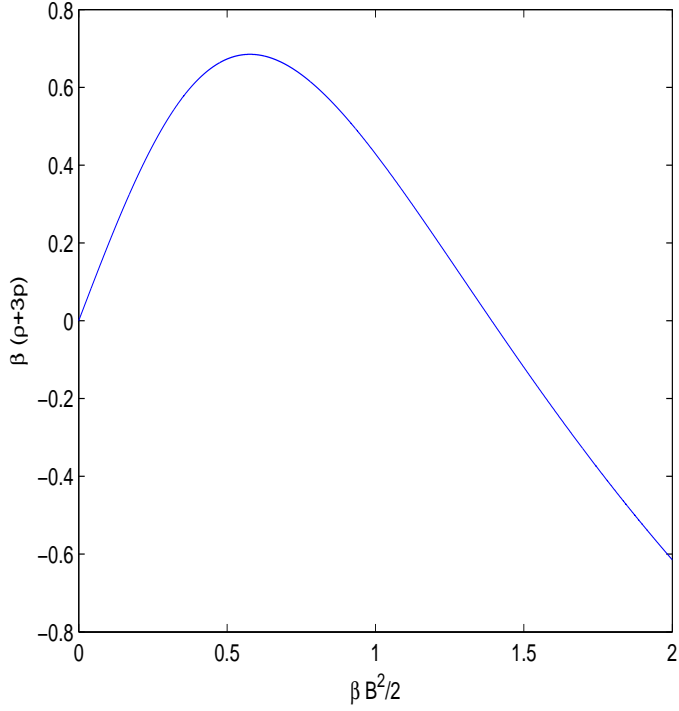}
\caption{\label{fig.1}The function  $\beta(\rho+3p)$ vs. $\beta B^2/2$. }
\end{figure}
By numerical calculations we find that for accelerating universe, $\rho + 3p < 0$, the inequality $\beta B^2/2>1.39175$ is needed (see Fig. 1). Thus, after Big Bang when strong magnetic fields take place, this requirement can be satisfied, and the magnetic field drives the universe to accelerate.
For FRW metric (14) from the energy-momentum tensor conservation, $\nabla^\mu T_{\mu\nu}=0$, the relation holds
\begin{equation}
\dot{\rho}+3\frac{\dot{a}}{a}\left(\rho+p\right)=0.
\label{20}
\end{equation}
For the case $\textbf{E} = 0$, from Eqs. (16),(17),  we obtain
\begin{equation}
\rho=\frac{1}{\beta}\arctan\left(\frac{\beta B^2}{2}\right),~~~~\rho+p=\frac{2B^2}{3\left[1+(\beta^2 B^4)/4\right]}.
\label{21}
\end{equation}
Taking into account Eqs. (21), and integrating Eq. (20), one finds the solution
\begin{equation}
B(t)=\frac{B_0}{a^2(t)}.
\label{22}
\end{equation}
According to Eq. (22), the magnetic field decreases when the scale factor increases due to inflation. The evolution of the energy density and pressure (at $\textbf{E}=0$) with the scale factor are given by
\begin{equation}
\rho(t)=\frac{1}{\beta}\arctan\left(\frac{\beta B_0^2}{2a^4(t)}\right),~p(t)=\frac{8B_0^2a^4(t)}{3\left[4a^8(t)+\beta^2B_0^4\right]}-\frac{1}{\beta}\arctan\left(\frac{\beta B_0^2}{2a^4(t)}\right).
\label{23}
\end{equation}
From Eq. (23) we obtain the limits
\begin{equation}
\lim_{a(t)\rightarrow 0}\rho(t)=\frac{\pi}{2\beta},~~\lim_{a(t)\rightarrow 0}p(t)=-\frac{\pi}{2\beta},~~\lim_{a(t)\rightarrow \infty}\rho(t)=\lim_{a(t)\rightarrow \infty}p(t)=0.
\label{24}
\end{equation}
Eqs. (24) show that at $a=0$ (at the beginning of the universe evolution) we have $\rho=-p$, i.e. the property of dark energy.
As a result, there are not singularities of the energy density and pressure at $a(t)\rightarrow 0$ and $a(t)\rightarrow \infty$.
This is an attractive feature of NLED model (1) proposed. The plot of the equation of state (EoS) $w=p(t)/\rho(t)$ versus $x=[2/(\beta B_0^2)]^{1/4}a(t)$ is given in Fig. 2.
\begin{figure}[h]
\includegraphics[height=4.0in,width=4.0in]{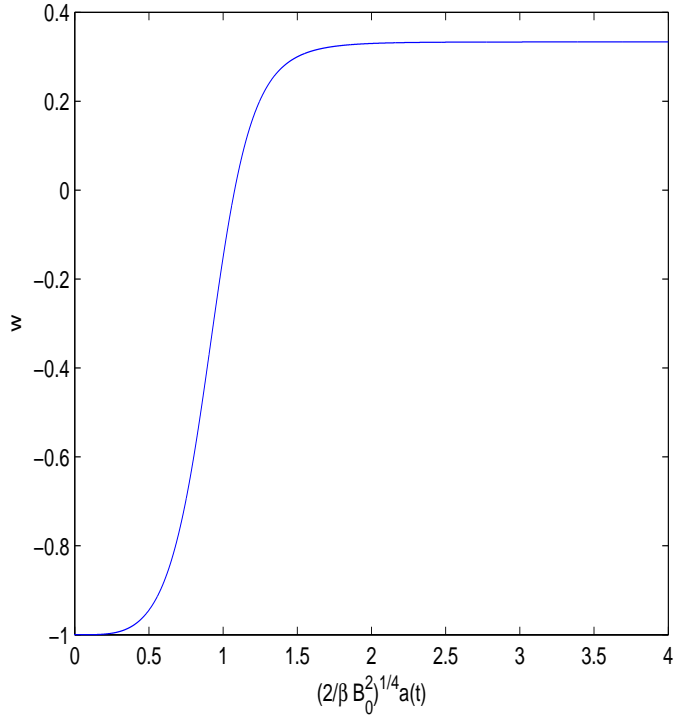}
\caption{\label{fig.2}The function  $w$ vs. $[2/(\beta B_0^2)]^{1/4}a$.}
\end{figure}
We obtain from Eqs. (23)
\[
\lim_{x\rightarrow\infty} w=\frac{1}{3}.
\]
Thus, at $a(t)\rightarrow \infty$ we have the ordinary EoS for ultra-relativistic case \cite{Landau}.
The curvature can be obtained from Einstein's equation (12) and the energy-momentum tensor trace (5),
\begin{equation}
R=\kappa^2T_{\mu}^{~\mu}=4\kappa^2\left[\frac{1}{\beta}\arctan(\beta{\cal F})- \frac{{\cal F}}{1+\left(\beta {\cal F}\right)^2}\right].
\label{25}
\end{equation}
Taking into consideration Eq. (22) we find the Ricci scalar depending on scale factor
\begin{equation}
R(t)=\frac{4\kappa^2}{\beta}\left[\arctan\left(\frac{\beta B_0^2}{2a^4(t)}\right)- \frac{2\beta B_0^2a^4(t)}{4a(t)^8+\beta^2 B_0^4}\right].
\label{26}
\end{equation}
The plot of the function $\beta R/\kappa^2$ versus $[2/(\beta B_0^2)]^{1/4}a$ is represented in Fig. 3.
\begin{figure}[h]
\includegraphics[height=4.0in,width=4.0in]{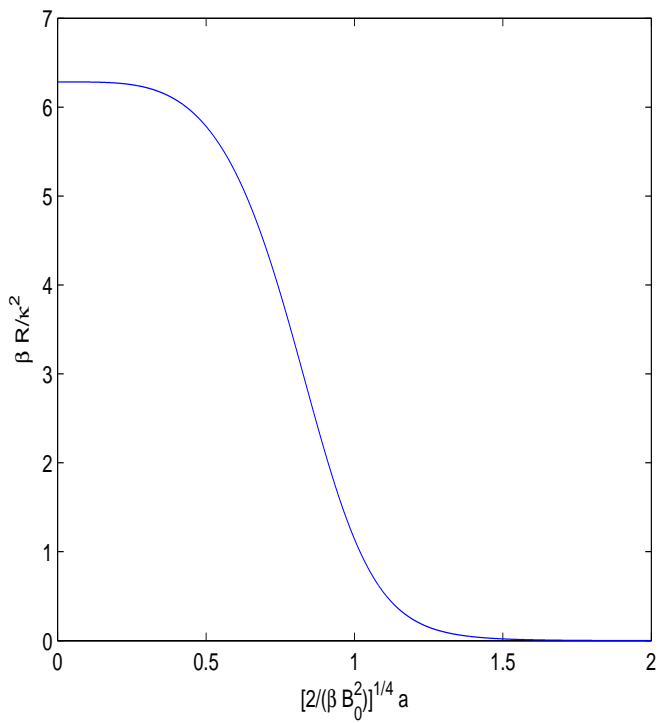}
\caption{\label{fig.3}The function  $\beta R/\kappa^2$ vs. $[2/(\beta B_0^2)]^{1/4}a$. }
\end{figure}
From Eq. (26) one obtains
\begin{equation}
\lim_{a(t)\rightarrow 0}R(t)=\frac{2\pi\kappa^2}{\beta},~~~~\lim_{a(t)\rightarrow \infty}R(t)=0,
\label{27}
\end{equation}
so that there are no singularities of the Ricci scalar. One can verify that the Ricci tensor squared $R_{\mu\nu}R^{\mu\nu}$ and the Kretschmann scalar $R_{\mu\nu\alpha\beta}R^{\mu\nu\alpha\beta}$ also have not singularities at $a(t)\rightarrow 0$ and $a(t)\rightarrow \infty$ because they are linear combinations of $\kappa^4\rho^2$, $\kappa^4\rho p$, and $\kappa^4p^2$ which, according to Eqs. (24), are finite.

Thus, at $t\rightarrow\infty$ there is no singularity,
the scale factor increases and spacetime becomes Minkowski spacetime (flat). We find from Eqs. (19),(22) that the universe accelerates at $a(t)<a_c(t)$, where the critical scale factor is $a_c(t)=0.7742\beta^{1/4}\sqrt{B_0}$. As a result, the universe inflation is described in the model suggested.

\section{The universe evolution}

Now we find the dependance of the scale factor on time. The second Friedmann equation for three dimensional flat universe is given by
\begin{equation}
\left(\frac{\dot{a}}{a}\right)^2=\frac{\kappa^2\rho}{3}.
\label{28}
\end{equation}
Taking into consideration Eq. (23), Eq. (28) becomes
\begin{equation}
\dot{a}^2 =\frac{\kappa^2a^2}{3\beta}\arctan\left(\frac{\beta B_0^2}{2a^4}\right).
\label{29}
\end{equation}
The solution to Eq. (29) can be represented in the form of the integral
\begin{equation}
t=\frac{\sqrt{3\beta}}{\kappa}\int \frac{da}{a\sqrt{\arctan \left(\beta B_0^2/(2a^4)\right)}}.
\label{30}
\end{equation}
For $a(t)<a_c(t)$ at $\beta B_0^2/(2a^4)\gg 1$ we obtain from Eq. (30) the approximate solution
\begin{equation}
a(t)=a_0 \exp \left(\frac{\sqrt{\pi}\kappa t}{\sqrt{6\beta}}\right).
\label{31}
\end{equation}
Eq. (31) describes the phase of inflation and corresponds to a de Sitter spacetime. Thus, without the cosmological constant and dark energy the model explains the early time universe acceleration.
For big values of the scale factor $a$ at $\beta B_0^2/(2a^4)\ll 1$ using the Taylor expansion the approximate value of the solution to Eq. (30) becomes
\[
\frac{\kappa t}{\sqrt{3\beta}}=\frac{a^2}{B_0\sqrt{2\beta}}-\frac{1}{36}\left(\frac{\beta B_0^2 }{2a^4}\right)^{3/2}+\frac{1}{240}\left(\frac{\beta B_0^2 }{2a^4}\right)^{7/2}
\]
\begin{equation}
-\frac{499}{332640}\left(\frac{\beta B_0^2 }{2a^4}\right)^{11/2}+{\cal O}\left(a^{-26}\right)+C.
\label{32}
\end{equation}
where $C$ is a constant of integration. The scale factor increases in the time. For $a(t)\gg a_c(t)$ and taking only the first term in the right side of Eq. (32) we obtain the approximate solution for late time of the universe evolution
\begin{equation}
a(t)=\left(\frac{2}{3}\right)^{1/4}\sqrt{\kappa B_0(t-t_0)}.
\label{33}
\end{equation}
The solution (32) for large time $t$ is $a\propto \sqrt{t}$ corresponding to the radiation era.
Thus, we have the same dependance of the scale factor as for the radiation era in Maxwell's theory.
For small cosmic time, in the early universe evolution, nonlinear corrections to Maxwell's theory are essential
and the de Sitter phase is realized. There is no singularity at the cosmic time $t=0$ and at $a_0< a_c$ the universe undergos the acceleration. After, at $a(t)>a_c(t)$ the the universe decelerates. The plot of the function  $y=\beta^{3/4}\ddot{a}/(\kappa^2\sqrt{B_0})$ vs. $x=[2/(\beta B_0^2)]^{1/4}a$ is presented in Fig. 4.
\begin{figure}[h]
\includegraphics[height=4.0in,width=4.0in]{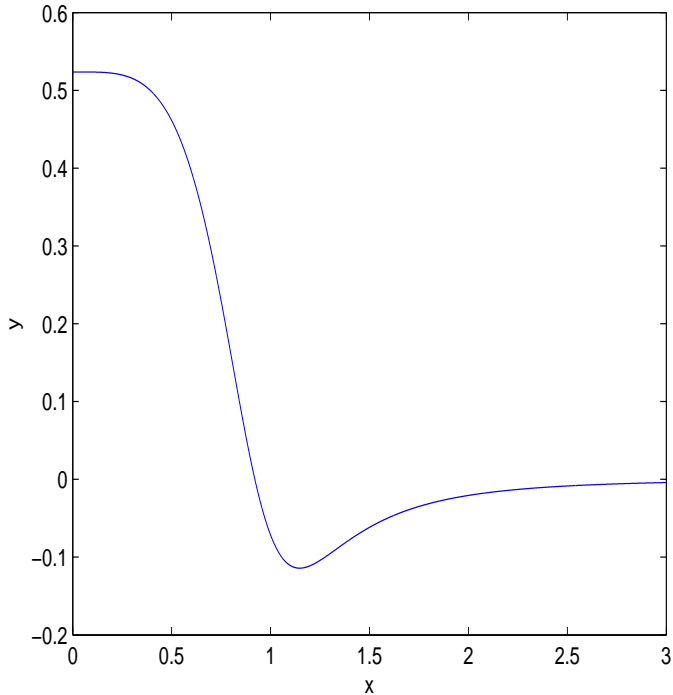}
\caption{\label{fig.4}The function  $y=\beta^{3/4}\ddot{a}/(\kappa^2\sqrt{B_0})$ vs. $x=[2/(\beta B_0^2)]^{1/4}a$.}
\end{figure}
Thus, the model describes inflation without singularities at the early epoch.

\subsection{Speed of Sound and Causality}

The speed of the sound should be less that the local light speed, $c_s\leq 1$ \cite{Quiros}. In this case the causality takes place. Another requirement is that the square sound speed is positive, i.e. $c^2_s> 0$, and then there is a classical stability. We find at $E=0$ from Eqs. (16), (17) sound speed squared
\begin{equation}
c^2_s=\frac{dp}{d\rho}=\frac{dp/d{\cal F}}{d\rho/d{\cal F}}=\frac{1-7(\beta {\cal F})^2}{3\left[(\beta {\cal F})^2+1\right]}.
\label{34}
\end{equation}
A classical stability occurs at $c^2_s> 0$,
\begin{equation}
1-\frac{7\beta^2 B_0^4}{4a^8(t)}> 0.
\label{35}
\end{equation}
The scale factor should obey the bound $a(t)>(\sqrt{7}\beta/2)^{1/4}\sqrt{B_0}\approx1.07\beta^{1/4}\sqrt{B_0}$ to have a classical stability. At this value of the scale factor the universe decelerates as the acceleration finished at $a_c(t)=0.7742\beta^{1/4}\sqrt{B_0}$. The inequality $c_s\leq 1$ holds for $a(t)>(\sqrt{7}\beta/2)^{1/4}\sqrt{B_0}$. As a result, the cosmological model admits subluminal fluctuations and the required bound $c_s\leq 1$ takes place at the deceleration phase. The plot of the function $c_s^2$ vs. $[2/(\beta B_0^2)]^{1/4}a$ is represented in Fig. 5.
\begin{figure}[h]
\includegraphics[height=4.0in,width=4.0in]{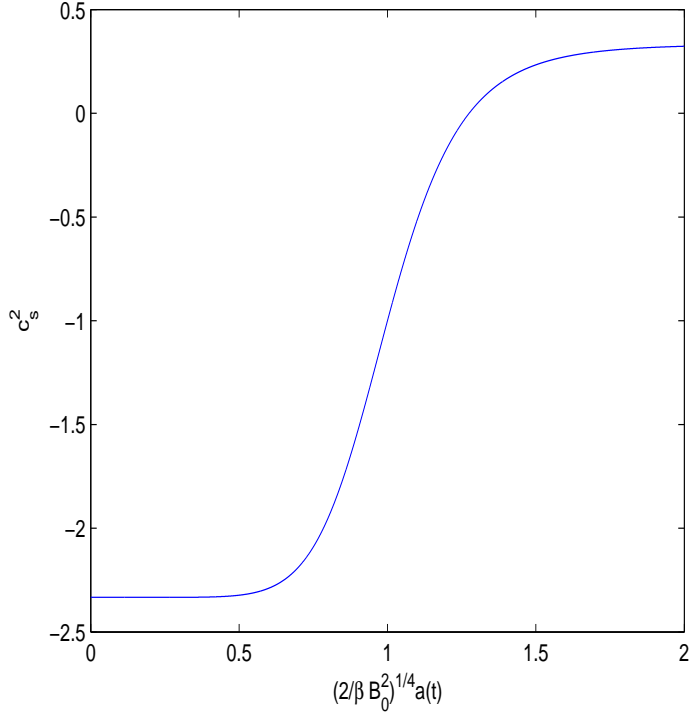}
\caption{\label{fig.5}The function $c_s^2$ vs. $[2/(\beta B_0^2)]^{1/4}a$.}
\end{figure}
A classical instability lasts when $a(t)<(\sqrt{7}\beta/2)^{1/4}\sqrt{B_0}$. Then at $a(t)>(\sqrt{7}\beta/2)^{1/4}\sqrt{B_0}$ the energy density perturbations do not grow. It should be noted that the inflationary period, i.e. accelerated expansion, ends up before the model reaches the stability regime according to the value of the squared speed of sound. So when the model is stable, then it decelerates.

\section{Cosmological parameters}

From Eqs. (21) we find that $\beta B^2=2\tan(\rho\beta)$. Then one obtains from Eqs. (21) the equation as follows:
\begin{equation}
p=-\rho+f(\rho),~~~~f(\rho)=\frac{2}{3\beta}\sin(2\rho\beta).
\label{36}
\end{equation}
Eq. (36) corresponds to EoS for the perfect fluid and shows that the pressure oscillates around the EoS $p=-\rho$. The plot of the function $p\beta$ versus $\rho\beta$ for the ranges $1\geq\rho\beta\geq 0$ and $15\geq\rho\beta\geq 0$ are given in Fig. 6 and 7, respectively.
\begin{figure}[h]
\includegraphics[height=4.0in,width=4.0in]{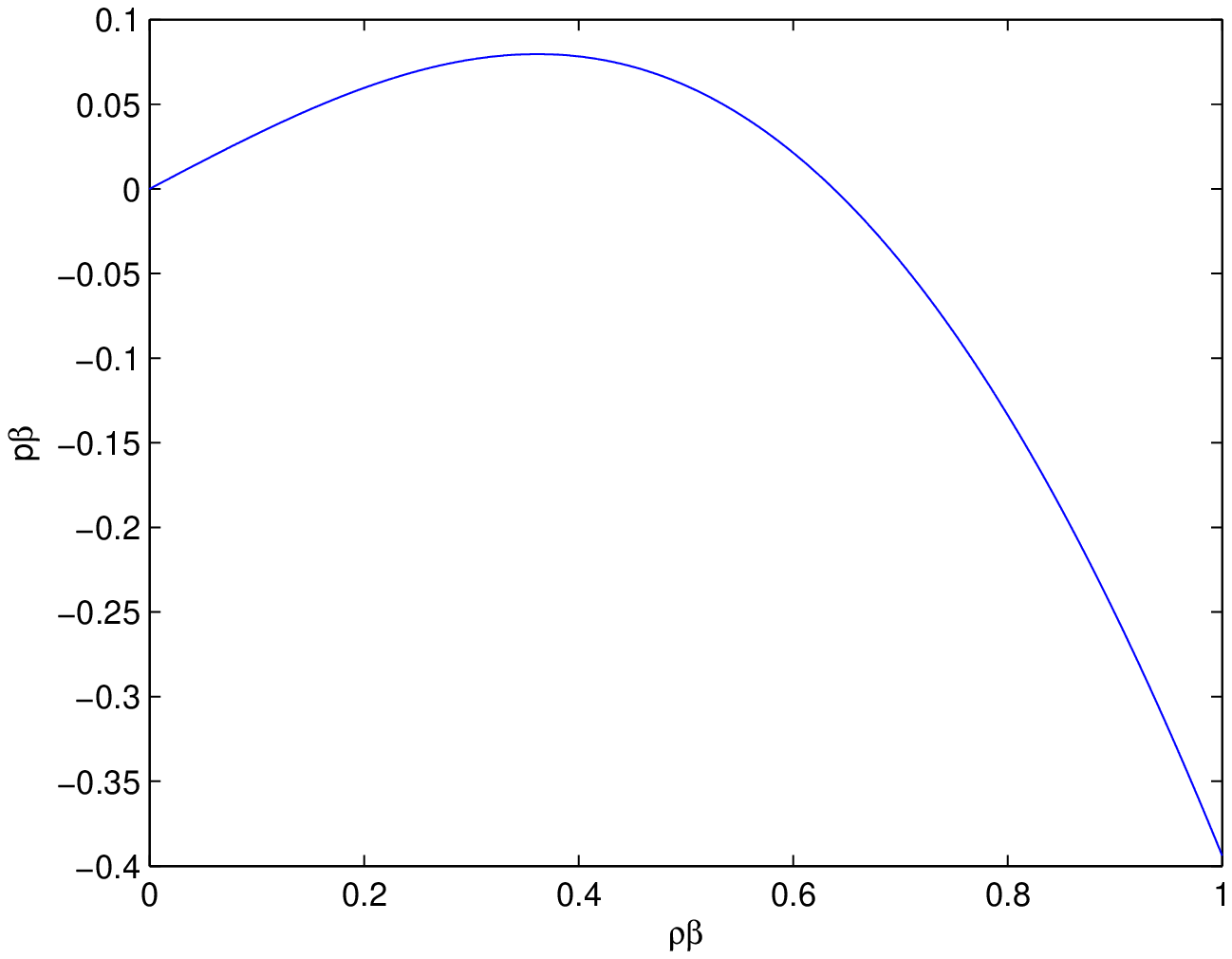}
\caption{\label{fig.6}The function $p\beta$ vs. $\rho\beta$.}
\end{figure}
\begin{figure}[h]
\includegraphics[height=4.0in,width=4.0in]{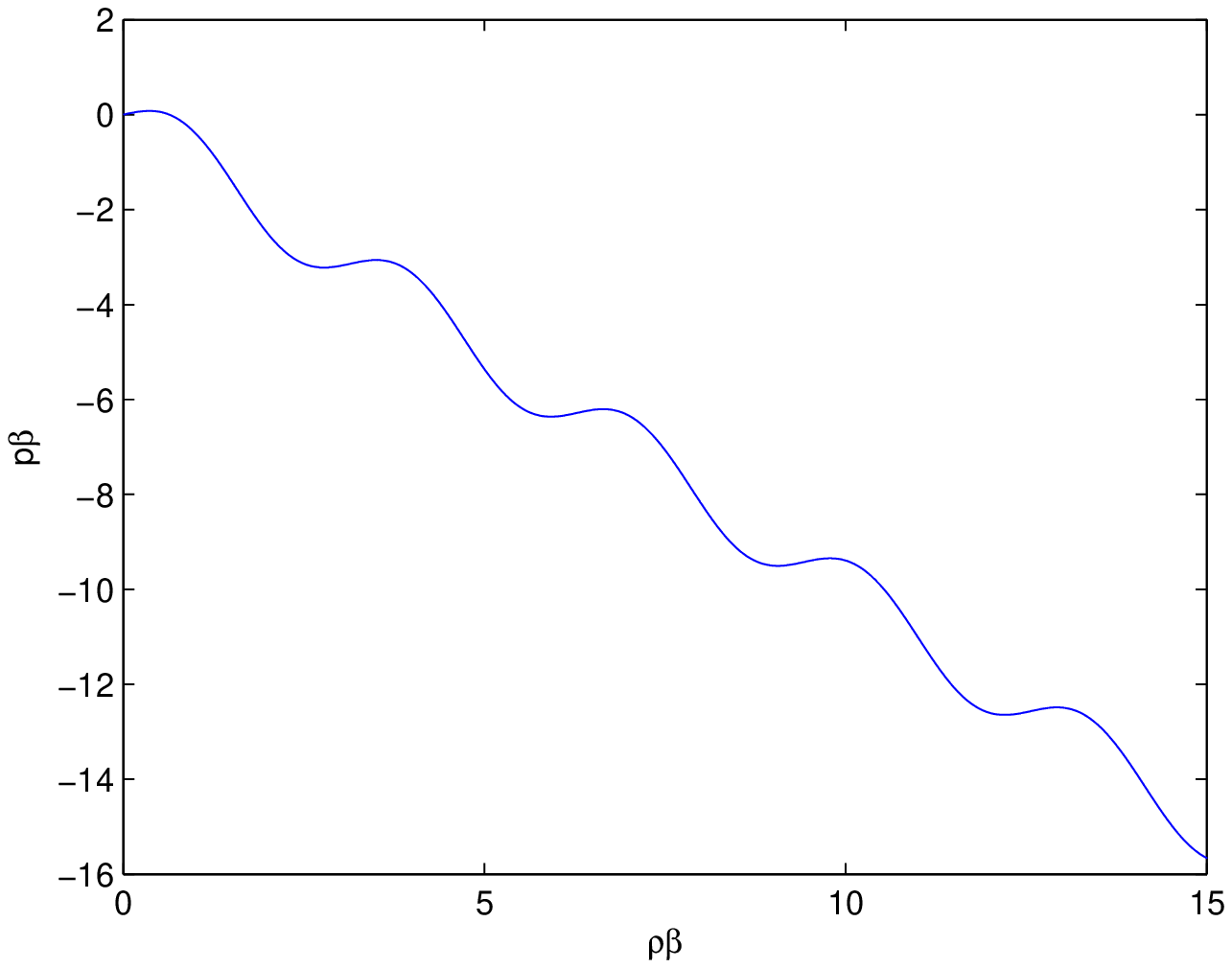}
\caption{\label{fig.7}The function $p\beta$ vs. $\rho\beta$.}
\end{figure}
From Eq. (36) we find that the pressure equals $0$ not only for $\rho=0$ but also for $\rho\beta=0.63785$ (see Fig. 6). The energy density decreases in time and after the time corresponding to the value $\rho=0.63785/\beta$ the pressure increases till the maximum $\beta p\approx 0.07959$ at $\rho\beta=0.36137$ and then decreases becoming zero. By the virtue of Eq. (36) we obtain EoS parameter
\begin{equation}
w=\frac{p}{\rho}=-1+\frac{2\sin(2\rho\beta)}{3\rho\beta}.
\label{37}
\end{equation}
It follows from Eq. (37) that $\lim w_{\rho\rightarrow \infty}=-1$ and $\lim w_{\rho\rightarrow 0}=1/3$ in accordance with Fig. 3. From the conservation of the energy, Eq. (20), and Eq. (36) one obtains
\begin{equation}
a=a_0\exp \left(-\int \frac{d\rho}{3f(\rho)}\right)
\label{38}
\end{equation}
and after the integration
\begin{equation}
a=a_0 \left(\frac{\cos(2\rho\beta)+1}{\sin(2\rho\beta)}\right)^{1/4}.
\label{39}
\end{equation}
Eq (39) also follows from Eq. (23) for the energy density at $a_0=(\beta B_0^2)^{1/4}$. If the condition $|f(\rho)/\rho\ll 1|$ is satisfied during the inflation the expressions for the spectral index $n_s$, the tensor-to-scalar ratio $r$, and the running of the spectral index $\alpha_s=dn_s/d\ln k$ are given by \cite{Odintsov}
\begin{equation}
n_s\approx 1-6\frac{f(\rho)}{\rho},~~~r\approx 24\frac{f(\rho)}{\rho},~~~\alpha_s\approx -9\left(\frac{f(\rho)}{\rho}\right)^2.
\label{40}
\end{equation}
As it was mentioned, at $\beta B^2/2>1.39175$ the universe undergoes the acceleration. This value corresponds to $\rho\beta\approx 0.9477$. Thus, at $\rho\beta> 0.9477$ the acceleration of the universe occurs. We note that at $\rho\beta>2.7$ the condition $|f(\rho)/\rho|< 0.2$ holds, and at $\rho\beta>5.8$ we have $|f(\rho)/\rho|< 0.1$. As a result, the parameters (40) can be fulfilled in the inflation phase. From Eqs. (40),(36) we obtain the relations as follows:
\begin{equation}
r=4(1-n_s)=8\sqrt{-\alpha_s}=\frac{16\sin(2\rho\beta)}{\rho\beta}.
\label{41}
\end{equation}
The PLANCK experiment \cite{Ade} and WMAP data \cite{Komatsu}, \cite{Hinshaw} give the results
\[
n_s=0.9603\pm 0.0073 ~(68\% CL),~~~r<0.11 ~(95\%CL),
\]
\begin{equation}
\alpha_s=-0.0134\pm0.0090 ~(68\% CL).
\label{42}
\end{equation}
But the BICEP2 experiment \cite{Ade1} gave for the tensor-to-scalar ratio the value $r=0.20^{+0.07}_{-0.05}~(68\% CL)$. It should be mentioned that the validity of this value was challenged.
If one excepts the value $r=0.13$ we get from Eqs. (41) the reasonable values for the spectral index $n_s=0.9675$ and the running of the spectral index $\alpha_s=-2.64\times 10^{-4}$. From Eq. (41) we find many solutions
for $\rho\beta$ that correspond to given $r$. We only mention the lowest values: $\rho\beta=3.12888,~6.30882,~9.38664$. The biggest values of $r$ are not written down. All these values correspond to the inflation phase.

\section{Conclusion}

We have introduced a new model of nonlinear electromagnetic fields that is the source of the gravitation field. The model has the dimensional parameter $\beta$ so that the scale invariance in this model is broken and the trace of the energy-momentum tensor does not vanish. The magnetic universe possessing a stochastic background with $<B^2>\neq 0$ was considered.
We show that the model for homogeneous and isotropic cosmology describes inflation and the universe accelerates at $a(t)<a_c(t)$. This phase corresponds to the de Sitter spacetime. The absence of singularities at the beginning of the universe creation in the energy density, pressure, and the Ricci scalar was demonstrated. In this model the magnetic field is the source of the universe acceleration. After the acceleration at $a(t)>a_c(t)$, the universe decelerates and we have the dependance of the scale factor $a(t)\propto \sqrt{t}$ corresponding to the radiation era.
The classical stability takes place at the deceleration phase at $a(t)>(\sqrt{7}\beta/2)^{1/4}\sqrt{B_0}$. We have demonstrated that the causality holds and the speed of the sound is less than the local light speed at the deceleration phase. Thus, a model of NLED presented allows the universe to accelerate due to a stochastic magnetic background at the early epoch. In the framework of inflationary cosmology we have described the universe inflation without introduction of the cosmological constant, dark energy and modification of GR.
We have calculated the spectral index, the tensor-to-scalar ratio, and the running of the spectral index that are in approximate agreement with the PLANK, WMAP, and BICEP2 data.
The model of NLED proposed makes the cosmological model be stable in the deceleration phase and it is questionable to use the model for describing early universe.

\end{document}